\documentclass[fleqn,usenatbib]{mnras}

\usepackage[T1]{fontenc}
\DeclareRobustCommand{\VAN}[3]{#2}
\let\VANthebibliography\thebibliography
\def\thebibliography{\DeclareRobustCommand{\VAN}[3]{##3}\VANthebibliography}

\usepackage{graphicx}	
\usepackage{amsmath}	
\usepackage{amssymb}	

\usepackage{newtxtext,newtxmath}

\title[IR variability of young solar analogs in M8]{Infrared variability of young solar analogs in the Lagoon Nebula}

\author[C. Ordenes-Huanca et al.]{C. Ordenes-Huanca,$^{1,2,3}$\thanks{E-mail: ccordenes@uc.cl}
M. Zoccali,$^{1,2}$
A. Bayo,$^{3,4,5}$
J. Cuadra,$^{3,6}$
R. Contreras Ramos,$^{1,2}$
L. A. Hillenbrand,$^{7}$
\newauthor
I. Lacerna,$^{2,8}$
S. Abarzua,$^{9}$
C. Avendaño,$^{9}$
P. Diaz,$^{9}$
I. Fernandez$^{9}$
and G. Lara$^{9}$ 
\\
\\
$^{1}$Instituto de Astrofísica, Pontificia Universidad Católica de Chile, Casilla 306, Santiago 22, Chile\\
$^{2}$Millenium Institute of Astrophysics (MAS), Nuncio Monseñor Sótero Sanz 100, Providencia, Santiago Chile\\
$^{3}$N\'ucleo Milenio de Formaci\'on Planetaria - NPF, Chile\\
$^{4}$European Southern Observatory, Karl Schwarzschild-Stra\ss e 2, D-85748 Garching bei M\"unchen, Germany \\
$^{5}$Instituto de F\'isica y Astronom\'ia, Facultad de Ciencias, Universidad de Valpara\'iso, Av. Gran Breta\~na 1111, Valpara\'iso, Chile\\
$^{6}$Departamento de Ciencias, Facultad de Artes Liberales, Universidad Adolfo Ib\'a\~nez, Av. Padre Hurtado 750, Vi\~na del Mar, Chile \\
$^{7}$Department of Astronomy, California Institute of Technology, Pasadena, CA 91125, USA \\
$^{8}$Instituto de Astronom\'\i a y Ciencias Planetarias, Universidad de Atacama, Copayapu 485, Copiap\'o, Chile\\
$^{9}$Citizen Scientists, Chile 
}

\date{Accepted XXX. Received YYY; in original form ZZZ}

\pubyear{2022}

\begin{document}
\label{firstpage}
\pagerange{\pageref{firstpage}--\pageref{lastpage}}
\maketitle

\begin{abstract}
T Tauri stars are low-mass pre-main sequence stars that are intrinsically variable. Due to the intense magnetic fields they possess, they develop dark spots on their surface that, because of rotation, introduce a periodic variation of brightness. In addition, the presence of surrounding disks could generate flux variations by variable extinction or accretion. Both can lead to a brightness decrease or increase, respectively. Here, we have compiled a catalog of light curves for $379$ T Tauri stars in the Lagoon Nebula (M8) region, using VVVX survey data in the $K_{s}$-band. All these stars were already classified as pre-MS stars based on other indicators. The data presented here are spread over a period of about eight years, which gives us a unique follow-up time for these sources at this wavelength. The light curves were classified according to their degree of periodicity and asymmetry, to constrain the physical processes responsible for their variation. Periods were compared with the ones found in literature, on a much shorter baseline. This allowed us to prove that for $126$ stars, the magnetically active regions remain stable for several years. Besides, our near-IR data were compared with the optical Kepler/K2 light curves, when available, giving us a better understanding of the mechanisms responsible for the brightness variations observed and how they manifest at different bands. We found that the periodicity in both bands is in fairly good agreement, but the asymmetry will depend on the amplitude of the bursts or dips events and the observation cadence.
\end{abstract}

\begin{keywords}
stars: formation -- stars: pre-main sequence -- stars: variables: T Tauri
\end{keywords}



\section{Introduction}
\label{sec:intro}
T Tauri stars are very young, recently formed low-mass stars with ages ranging from $1$ to $10$ Myr. These objects are still contracting, surrounded by disks from which, in many cases, they are still accreting. Despite the possible attenuation of these disks (when present), T Tauri stars represent a group of Young Stellar Objects (YSOs) evolved enough to be detectable in the optical \citep{Bertout1989}.

Accretion will determine one of the classifications of this type of stars. The ones that accrete material from their disks are known as classical T Tauri stars (CTTSs), whereas the ones that do not have a disk or, if they do, are not accreting, are called weak-line T Tauri stars (WTTSs). They can be discriminated using the strength of their H$\alpha$ emission line. In the presence of accretion, as in a CTTS, the equivalent width of the H$\alpha$ line will be larger than in the absence of this process, as in a WTTS \citep{Herbst2012}. Furthermore, it has been shown that the width of the H$\alpha$ line at $10 \%$ of its peak intensity allows to divide accretors from non-accretors at a limit of 200 km s$^{-1}$ \citep{Natta_2004}. However, due to chromospheric emission of late-type stars, the limit for determining whether a T Tauri star is accreting or not will depend also on its spectral type \citep{White_2003, BarradoNavascues_2003}.

One of the main features of the emission that we observe for these objects is that they are intrinsically variable. They have strong magnetic fields that produce dark, cold spots in their stellar photosphere. Coupled with rotation, spots introduce a periodic and nearly sinusoidal flux variation with periods in the range from $1$ to $10$ days and amplitudes around $\sim 0.1$ mag in the near-IR \citep{Wolk_2013, carpenter2001, Carpenter_2002}. For stars that are accreting material, their brightness variation is rather random since accretion is, by nature, a stochastic process. However, in several cases variation due to this mechanism could be periodic because of rotationally modulated hot spots located in the region where disk material hits the surface of the star \citep{Herbst_1994, Bouvier_2020}. The temperature difference between a hot spot and the stellar photosphere is higher than the one for cool spots. Consequently, they induce flux changes with magnitudes higher that the ones for cool spots, about $\sim 0.6$ mag in the near-IR \citep{Wolk_2013}. 

The effect of spots in the emission of these stars is more evident at short wavelengths, although they can also be observed at longer wavelengths, such as the near-IR. When periodic light curves are related to spots, this give us in turn, a measurement of the rotation period of the star \citep{Rydgren_1983}. Therefore, they can help us understand the evolution of the angular momentum in the star's road to the MS. Accretion in disk-bearing stars is usually associated with lower rotation rates compared to stars that do not have a disk due to the so-called disk-locking effect \citep{Koenigl_1991}. In these systems, the spin down torques conveyed by magnetic field lines that connect the star to slower disk regions are balanced by the spin up ones on disk regions that have angular speeds higher than the star, preventing the decrease of the rotation rate. Therefore, having the rotation periods of stars with and without a surrounding disk allows us to confirm if the angular momentum is regulated by the presence of such structure.

On the other hand, for disk-bearing stars, variations at longer wavelengths are also affected by absorption by dust grains that build up this structure. Particularly, in the infrared, dust at different radii will reveal information about the disk inner parts (near-IR) or its more distant, outer parts (mid and far-IR). Variable extinction, associated with accretion columns or clumps in the disk will also introduce decreases in the observed flux of the star, either periodic or stochastic, when these structures cross our line-of-sight \citep{Kesseli_2016}. Flux dips, superimposed on a more or less quiescent level of brightness, can be observed in the light curves of these stars that are, therefore, called "dippers". For the ones with periodic behavior, their timescale ranges from days to weeks and their amplitudes are about $\sim 0.16$ to $1.5$ mag at J-band. The physical origin of the dips is related to the development of high latitude "warps", or enhanced density regions, in the disk, that pass through our line-of-sight. Both of them could block part of the emission of the star, creating dips in the light curves \citep{Morales_Calder_n_2011, Bodman_2017}. 

However, accretion can be also responsible for increases in the stellar emission. When material from the disk impacts the surface of the star, shocks are developed producing the so called bursts. This introduces sudden increases in brightness. Stars with this behaviour are called "bursters", and their bursts are also superimposed on a quiescent level of emission. If accretion is stable, hot spots are going to be developed and this stability will determine the type of periodicity of the light curve \citep{Stauffer_2016}. In general, bursts are related with aperiodic behaviour of flux changes due to variable accretion and they can last from days to months, although they can also have durations of hours \citep{Hillenbrand_2015, Cody_2014}. However, there has been evidence of periodic outbursts due to accretion, when the latter is regulated by the presence of a low-mass companion \citep{Dunhill_2015, Teyssandier_2020, Guo_2022}. Most importantly, accretion processes will leave a signature in optical and infrared light curves that, depending on the view angle or the location of the spot, could be well correlated. Therefore, besides the T Tauri status of a given source, the presence or absence of a disk and its inclination in relation to the observer, also dictates the features of what we observe, including its type of variability. 

As light curves can have different morphologies according to the physical process from which the variability originates, \citet{Cody_2014} proposed two parameters that help us to classify a light curve according to their degree of periodicity (Q), that tell us if flux fluctuations have a stable period or not, and asymmetry (M), which indicates if the flux has a tendency or not to decrease or increase. These two criteria can be used as a guideline to identify the process that dominates the brightness variation in a given light curve.

This study is focused on the $K_{s}$-band variability of T Tauri stars located in the Lagoon Nebula (M8), an HII region in the Sagittarius arm of the Milky Way. Associated with it, there is the open cluster NGC6530, which includes thousands of pre-main sequence (PMS) stars and several O-type stars that ionize the M8 nebula \citep{Arias_2007}. Many studies have been made in this region regarding the physical parameters of these sources \citep{Damiani_2004, Kalari2015, Prisinzano2019} and also studying their variability \citep{Henderson_2012, Venuti_2021} at different wavelengths. However, the timespan for these last studies ranges from weeks to months. In our case, using approximately eight years of $K_{s}$ data from VISTA Variables in the V\'ia L\'actea \citep[VVV and its extension VVVX; hereafter simply VVVX]{minniti+2010} survey, we are able to observe changes over longer periods of time and classify the light curves guided by the $Q$ and $M$ parameters. This is one of the first studies which considers these metrics to classify VVVX light curves and one of the few that is based on ground-based data \citep{Bredall_2020, Hillenbrand_2022, Guo_2022}. 

At near-IR wavelengths, the presence of the disk can dominate the brightness changes, losing a periodic signal in several cases. Nonetheless, if hot/cool spots drive the flux variations, we can observe periodic light curves in the $K_{s}$-band. As mentioned, correlation between optical and near-IR light curves can occur for several variability mechanisms. 

In this work, we compared our $K_{s}$ light curves with the available optical ones. Particularly, we cross-correlated our catalog with the one of \citet{Venuti_2021} and, for the common sources, we investigated the correlation between the classification of the light curves, to confirm that the same physical mechanism dominates the variability in both bands.

The article is organized as follows. Section~\ref{sec:VSZ} details the motivation of the present work, related to the Variable Star Zoo citizen science experiment. In Sec.~\ref{sec:catalog} we describe how the catalog of light curves in the $K_{s}$-band for T Tauri stars in the Lagoon Nebula is compiled; the classification of the light curves according to $Q$ and $M$ parameters is discussed in Sec.~\ref{sec:QM}. A comparison between our results and previously found periods is presented in Sec.~\ref{sec:comparison_periods} and the correlation between our light curve classification with the one in the optical band is described in Sec.~\ref{sec:correlation_optical}. Finally, we conclude on Sec.~\ref{sec:conclusions}.\\\

\section{The Variable Star Zoo citizen science experiment}
\label{sec:VSZ}

The idea of the present work originated from a citizen science experiment that we conducted a few years ago. We briefly summarise it here, both as an acknowledgement to the users who invested their time to the classification of variables, and to share part of our experience with other citizen science project builders.

\begin{figure}
	\includegraphics[width=\columnwidth]{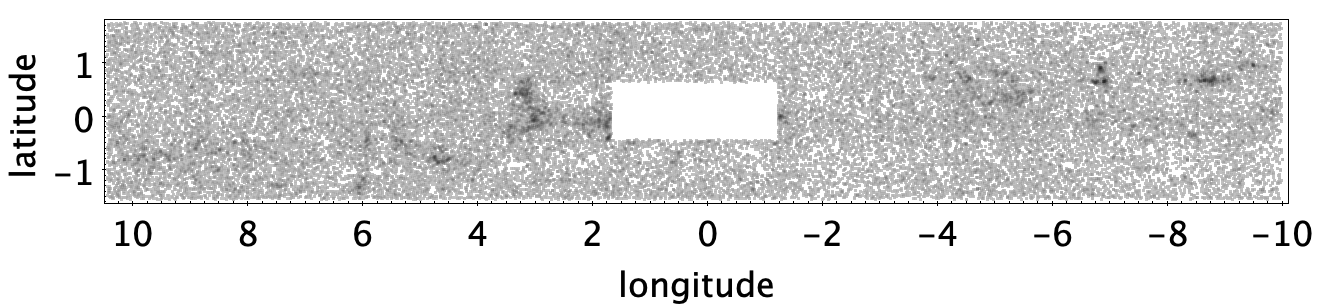}
    \caption{Galactic coordinates of the input catalogue of the Variable Star Zoo experiment. The innermost region was not included, as the photometry was noisier there, and the candidate light curves could not be easily compared with standard templates in near IR.}
    \label{fig:vsz}
\end{figure}

In 2018, as part of the outreach activities of the Millennium Institute of Astrophysics, we launched a citizen science experiment called Variable Star Zoo, hosted as an official project within the international platform Zooniverse\footnote{ https://www.zooniverse.org/projects/ilacerna/variable-star-zoo}. 
Variable Star Zoo had to pass an evaluation process before being part of Zooniverse. The project aimed to classify variable stars by looking at their phased light curve and comparing them with a list of templates. The input catalogue included almost 55,000 candidate variable stars, within $-10<l<10$ and $-1.6<b<1.8$ (Fig.~\ref{fig:vsz}), extracted from the VVVX multiepoch PSF-photometry catalogue available to our group \citep[derived as explained in][]{contrerasramos17}. The templates provided were the following: RR-Lyrae/Cepheid, Mira, eclipsing binary, microlensing event, unusual or just noise. The first class included sinusoidal light curves that were meant to be separated afterward, according to their period. The project was concluded about one year and a half later when at least 15 different citizen users had classified each candidate variable. More than 5000 users helped us complete more than 860,000 classifications. We also included an initial sample of $\sim$1500 well-known variables that we used to validate the reliability of the classifications of each user. The control sample increased with time by adding candidate variable stars with an agreement of over 75$\%$ in the classification made by 11 or more volunteers. We contacted about 130 enthusiastic volunteers that helped us with more than 500 classifications each and, at the same time, had a reliable performance in the classifications. In this way, we learnt that they connected mostly from Chile but also from, e.g. Europe, India, the US, Brazil, Mexico, Canada, and other countries.

In reviewing the results, we noticed a few overdensities of stars classified as RR-Lyrae/Cepheids, with periods indeed compatible with those of Cepheids, but clustered around known star forming regions such as M8. Their proper motions were also consistent with being members of this region. They could not be Cepheids, because at a distance of a few kpc, they would have been badly saturated in VVVX. Given their magnitudes and color, they could only be T Tauri stars, and indeed we quickly found out that many of them had been already identified and studied in the literature. No near infrared light curve was available for them, though. In fact, even for other T Tauri stars, no study was available, in the near infrared, with such a large time baseline, and this motivated us to carry on the present analysis.

Although, in the end, the catalogue discussed here was built independently from the results of the Variable Star Zoo experiment, we recognize that the latter had the important role to draw our attention to these variable stars. We acknowledge here the valuable time invested by all the citizen scientists in helping us with classification and motivating the present work. Given our public, national funding from the ANID Millennium Institute of Astrophysics (MAS), a few of the citizen scientists from Chile are co-authors of the present paper as a reward for having classified $>10,000$ light curves, each.

\section{Catalog compilation}
\label{sec:catalog}
\subsection{Preliminary list of NGC6530 and M8 T Tauri sources}
In order to compile a list of coordinates of T Tauri stars in the M8 region, three previous studies related to NGC6530 and the Lagoon Nebula were considered. One of them is the work by \citet{Henderson_2012}, where rotation periods for almost $300$ sources in NGC6530 are presented. The authors monitored the Lagoon Nebula using SMARTS at Cerro Tololo Inter-American Observatory (CTIO) for $35$ nights. Using these data, light curves in the Cousins I-band were produced and, consequently, rotation periods were determined. Their sample includes the T Tauri status for a few stars, from spectroscopy previously presented in \citet{Arias_2007} and \citet{Prisinzano_2007} in addition to previously identified spectroscopic binaries (SB). X-ray luminosity of many of them is also given and obtained from the work by \citet{Damiani_2004}. 

The second work used for the catalog composition is the one by \citet{Kalari2015}, where the authors derived accretion rates for more than $200$ CTTSs candidates in the Lagoon Nebula. These rates were computed considering $ugri$ H$\alpha$ photometry from VST Photometric H$\alpha$ survey+ (VPHAS) and selecting the sources that showed H$\alpha$ excess, an indicator of accretion. This excess was obtained by analyzing photometry in $u$ and H$\alpha$ bands. This study, therefore, is restricted only to CTTSs.

Finally, the last study included for the preparation of the VVVX catalog is the one of \citet{Prisinzano2019}. In this work, different physical parameters of T Tauri stars are obtained in order to derive the age distribution of NGC6530 sources. In their sample, they have more than $600$ confirmed members and their T Tauri classification includes WTTSp (WTTSs with only photospheric emission and no signs of IR excess) and CTTSe (CTTSs that do have IR excess). This designation was obtained using photometric data from \citet{Kalari2015} and spectroscopic data from the Gaia-ESO survey (GES), particularly the full width at zero intensity (FWZI) of the H$\alpha$ line. A summary of the stars studied in these three articles and the method to obtain the T Tauri status of them is presented in table \ref{tab:summary_articles}.\\

\begin{table}
\centering
	\caption{Number of stars studied in the three works considered to compile our catalog and the method used to obtain the T Tauri classification of the sources. For the one of \citet{Henderson_2012}, not all the sample had information about the classification.}
	\label{tab:summary_articles}
\resizebox{\columnwidth}{!}{%
\begin{tabular}{|c|c|c|c|c|c|}
\hline
Article & \# stars & Periods & \# CTTS & \# WTTS & TT status based on \\ \hline
Henderson et al. (2012) & 290 & Yes & 10 & 15 & spectroscopy \\ 
Kalari et al. (2015) & 235 & No & 235 & 0 & photometry \\ 
Prisinzano et al. (2019) & 661 & No & 333 & 328 & spectroscopy + photometry \\ \hline
\end{tabular}%
}
\end{table}

We cross-identified the data of these three studies with our VVVX data and a list of unique members with their positions was assembled. This procedure resulted in $942$ VVVX variable sources already included in one, or more, of the previous works. Every unique source from the literature, but one, had a counterpart on the $K_{s}-$band. For four stars, the light curves have less than four measurements, so that, we could not generate a proper light curve. Therefore, the initial catalogue for the present work includes a total of $937$ light curves in $K_s$.

\subsection{Period search}
\label{sec:period} 

For all the $937$ light curves, we applied the Lomb-Scargle period search algorithm \citep{Lomb1976, Scargle1982}, implemented in Astropy \citep{VanderPlas_2018}, to obtain their two most probable periods, namely $P_{1}$ and $P_{2}$. To accomplish this, we searched for peaks in the periodogram in the frequency range of $7.4 \times 10^{-4}$ $d^{-1}$ and $2.5$ $d^{-1}$. The lower limit is related to the baseline of VVVX observations, corresponding to almost half of the entire follow-up (about $1300$ days). The upper frequency limit was chosen due to the breakup velocity of a typical T Tauri star, about $0.4$ days \citep{Bertout1989}. However, it is important to mention that we refer to the peaks on the Lomb-Scargle periodogram as "periods", even though some of them may be more appropriately called variability timescales, as the light curves do not show strictly repeating waveforms.

Due to the cadence of VVVX observations, near $\sim 1$ day, $44$ light curves showed that their two most probable periods were around this value or values $0.5$ or $2$ days. Furthermore, when the window function for the observations was computed, it showed peaks at those values and, also, in $P \approx 365$ days. Consequently, those estimates could be artificial or false periods and, because of this, we visually inspected their light curves. Only the ones in which the folded light curve looked like a true variability pattern, having a sinusoidal variation with no gaps between points, were included in our sample. We looked for this type of patterns because at this timescale ($2$ days at most) they are more probable to originate a periodic variability. A total of $5$ sources were kept and analyzed as the rest of the sample, but the remaining ones were discarded from this study.

For the other sources, in order to select one of the two periods found, $P_{i}$, we first compare our periods $P(K_{s})$ with the ones presented in \citet{Henderson_2012}, namely $P(I)$. For the $274$ stars in common that have at least one true period, we kept the period closer to the one in the I-band. In $31$ cases, the closest period was an artificial period with no clear pattern of variation and with a difference of $|P(K_{s})-P(I)|>0.4$ days. These were also not analyzed in this study. Additionally, for two cases we modified the period searching limit to a lower value of $0.1$ days and obtained a period similar to the one I-band. We kept those values just because they were already found and confirmed with our data. 

For the rest of the sources, those not in \citet{Henderson_2012}, we kept the most probable period, $P_{1}$, only if it was far from an artificial value (a difference of at least $0.02$ days). Otherwise, we considered the second period, $P_{2}$, as the true one.

According to the filters mentioned above, we kept a list of $862$ light curves with true period values for the most probable one, and an additional $5$ that passed our visual inspection and were added to the $862$ stars. These gave us a total of $867$ light curves which are analysed further.

\subsection{Final catalog}

\begin{figure}
	\includegraphics[width=\columnwidth]{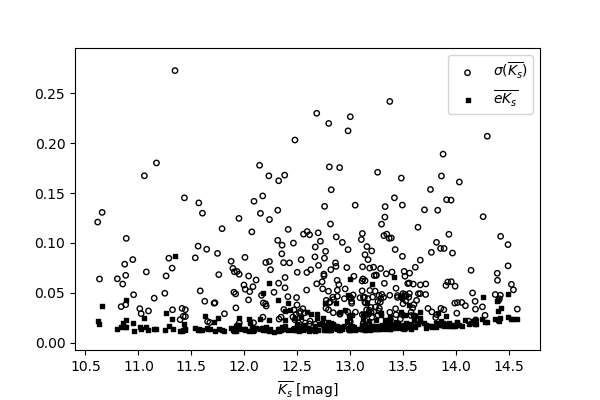}
    \caption{Standard deviation $\sigma(K_{s})$ against the mean magnitude $\overline{K_{s}}$ for each of our sources (open circles). Black points illustrate the mean photometric error vs. $\overline{K_{s}}$. This comparison indicates that the variability of our sources is above the noise level.}
    \label{fig:meanmag_dev}
\end{figure}

Lastly, two filters were applied to the list of sources with variable $K_s$ light curves. We kept the data that had at least $40$ data points and for which the median data error was less than 10$\%$ of the light curve amplitude, defined as $\Delta K_{s} = K_{s, max}-K_{s, min}$, in order to be able to measure a significant variation. The latter condition, however, was not required for stars that had periods very similar to the one in \citet{Henderson_2012}. How the standard deviation $\sigma(K_{s})$ relates to the mean magnitude $\overline{K_{s}}$ is shown in Fig.~\ref{fig:meanmag_dev} as open circles. In this plot is also shown the mean photometric error of these sources (black points). This remarks how the photometric error is kept at a low level compared to the variability of the sources in our catalog. 

Additionally, we kept only the stars with periods less than $40$ days, as this is the typical timescale of variability in this kind of sources \citep{Fischer_2022}, and also because the periodicity parameter (see Sec.~\ref{sec:QM}) do not give meaningful results for longer periods. Sources with P$>40^d$ will be discussed further below. With these conditions we were left with $355$ stars. Further, $18$ stars had mean magnitudes near the saturation limit and also had close companions in their VVVX images. In those cases, blending can significantly affect the magnitude measurement in a way that depends on seeing, and therefore it introduces time dependent variations that would affect our analysis. Therefore, these sources were removed from our sample. This gave us a catalog with $337$ T Tauri stars with periods less than $40$ days.

\begin{figure}
	\includegraphics[width=\columnwidth]{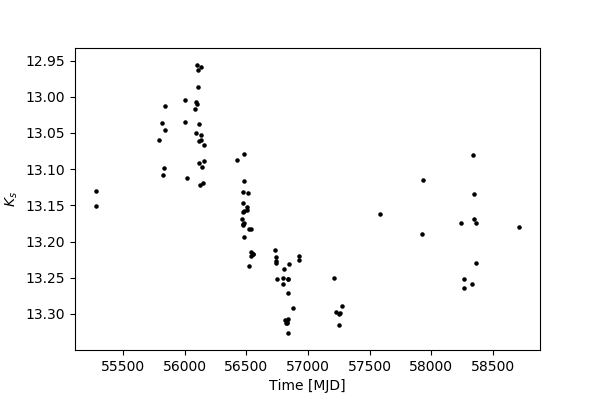}
    \caption{Example light curve of a source having a long timescale behaviour. A consistent decrease in brightness is observed, particularly in the seasons were the cadence is higher, $56000 < t< 57000$ MJD.}
    \label{fig:long_timescale}
\end{figure}

Sources with P$>40^d$ were visually inspected. $42$ stars showed a persistent increase or decrease in the flux, from one season to another. These were added to the catalog (see, e.g., Fig.~\ref{fig:long_timescale}). These stars do not show a periodic variability, but rather a consistent variation on the long time. Therefore, the periods derived with our algorithm do not have the usual meaning, but rather they indicate a characteristic timescale. As a double check, we repeated the period search with different time limits. For several stars, the original period was found. For the others, we can only say that they have flux changes in $P>40$ days. All the filters mentioned are listed in table \ref{tab:conditions}.

\begin{table}
	\centering
	\caption{Conditions or filters applied to the VVVX light curves with the number of stars that met the condition.}
	\label{tab:conditions}
	\begin{tabular}{cc} 
		\hline
		Condition & \# stars\\
		\hline
		Pre-MS candidates detected in VVVX & 937\\
		$P_{1}$ and $P_{2}$ $\neq$ false & 893\\
		$P_{i}$ $\neq$ false or $P_{i}$ $\approx$ false with sinusoidal pattern & 867\\
		$40$ data points, no saturation, true variability, $P \leq 40$ days & 355\\
		No close companions & 337\\
		$P \leq 40$ + Long timescale & 379\\
		\hline
	\end{tabular}
\end{table}

\begin{figure}
    \centering
	\includegraphics[width=\columnwidth]{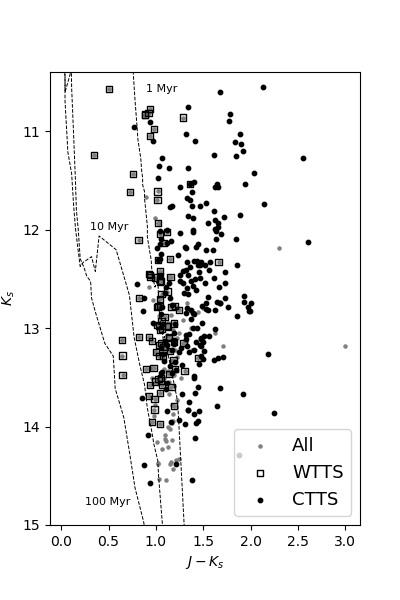}
    \caption{Color-magnitude diagram $K_{s}$ \emph{vs.} $J-K_{s}$ for the T Tauri stars in our catalog of $K_{s}$-band light curves (gray dots). Available T Tauri status is indicated by squares for WTTSs and by black dots for CTTS. The dashed lines are PMS isochrones from \citet{Siess_2000} at the indicated ages.}
    \label{fig:cmd}
\end{figure}

With all the selections above, the catalogue of T Tauri stars for which we have reliable $K_s$ light curves includes 379 sources. The timespan of the observations comprises data points from March 2010 to August 2019, but in some cases it extends from August 2011 to August 2019. The positions of stars in our catalog in the color-magnitude diagram (CMD) are shown in Fig.~\ref{fig:cmd} along with PMS isochrones from \citet{Siess_2000}, considering $A_{V}=1.1$ and distance of $1.25$ kpc \citep{Prisinzano_2005}. A a portion of these stars, particularly those that have been classified as CTTSs, are located in the red part of the diagram. Hence, they present an excess of emission in the $K_{s}$-band, related to the presence of a disk.

\begin{figure}
	\includegraphics[width=\columnwidth]{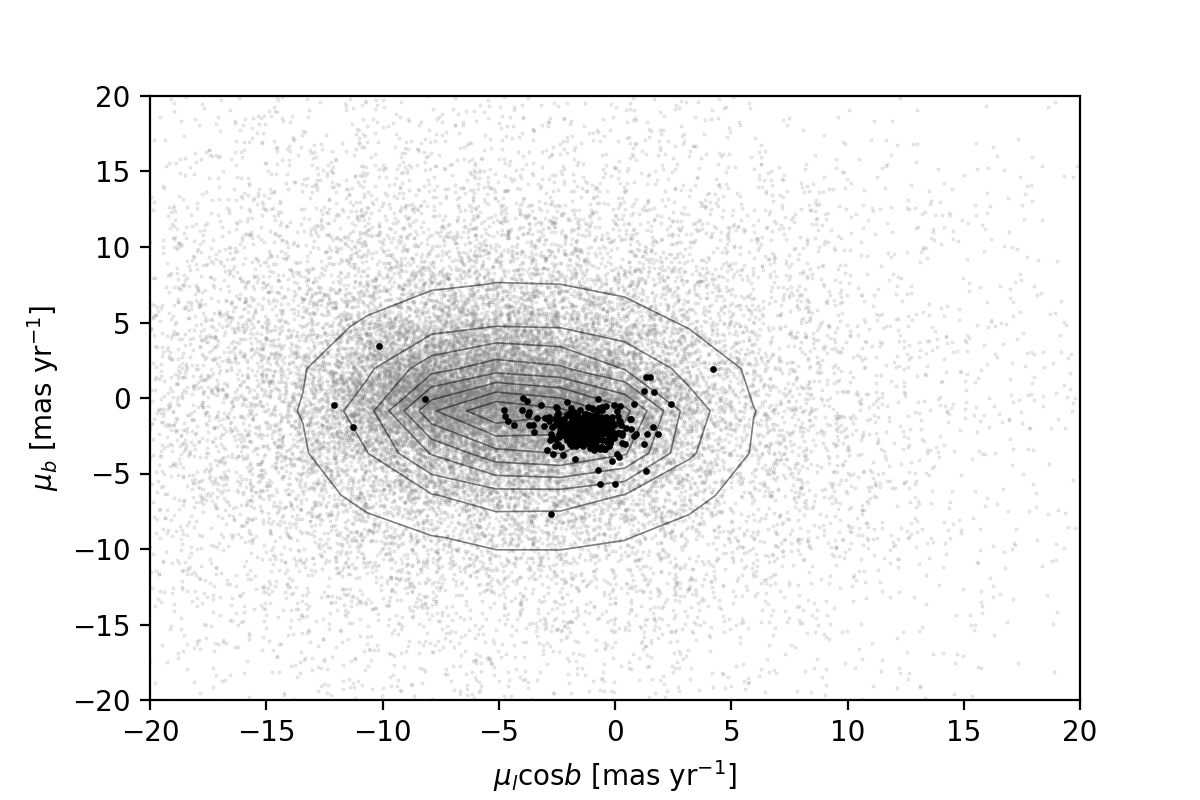}
    \caption{Vector point diagram for stars in our catalog (black points). Gray points represent all the stars with VVVX proper motions within a radius of $12$ arcmin from the center of NGC6530 and contours indicate density levels. Most of the sources analyzed here are clustered in a narrow region of the plot, indicating that they are true members of the cluster.}
    \label{fig:pms_vvv}
\end{figure}

To ensure that these stars are true members of the open cluster NGC6530 or the M8 region, the plot of their VVVX proper motions is shown in Fig.~\ref{fig:pms_vvv}. Black points represents the sources that compose the VVVX catalog of light curves. The plot indicates that the vast majority of sources fall in a narrow region confirming their coherent motion and membership. The ten stars whose proper motion differ by more than $3\sigma$ from the mean of the other were anyway kept because they have either parallaxes consistent with NGC6530 \citep{Gaia_edr3}, or they are listed as members of NGC6530 (or M8) in the Simbad database.

\section{Light curve classification}
\label{sec:QM}
Different physical mechanisms are responsible for the variability of T Tauri stars, as discussed in Sec.~\ref{sec:intro}. Each of them is manifested in the morphology of the light curves of these sources. Because of this, \citet{Cody_2014} proposed two parameters that allow one to quantify the degree of periodicity $Q$ and asymmetry $M$ of a given light curve. These allow us to classify the light curves in nine morphological categories, depending on the physical process that gives rise to the brightness variation. Those parameters have already been used in several optical and infrared studies \citep{Stauffer_2016, Cody_2018, Venuti_2021} leading to a good identification of the underlying process causing flux changes. However, these metrics were designed to work on space-based data, with high cadence and uniform sampling. The VVVX data presented here, although with larger time baseline, have lower cadence, uneven sampling and larger photometric error. Therefore, the limits and the definition of these parameters were slightly modified, keeping their meaning intact (see, for instance, \citet{Hillenbrand_2022}). The $Q$ and $M$ metrics are defined and described in \citet{Cody_2014}, nonetheless we will briefly recall them here. 

\subsection{Periodicity} 

The parameter $Q$ divides the light curves in three categories: periodic, quasi-periodic and aperiodic. The first two classes describe sources maintaining their periods constant over the entire observation time, but for quasi-periodic stars, the shape of the light curve is evolving with time. Aperiodic variables are, instead, those displaying no stable period or stochastic variations. To compute this metric, we first folded the light curves using the selected period and smoothed them using a boxcar method, considering a window size of $10 \%$ of the period. Then, we subtracted the smoothed curve from the raw data to obtain the residuals. Any data point above or below $3\sigma$ from the mean of those residuals was considered an outlier and was removed from the light curve. With the remaining points and to avoid edge effects, as in \citet{Hillenbrand_2022}, we folded the light curves again and repeated it for two more consecutive times, giving us a 3 times folded light curve. Then, we smoothed the outlier-free data, with the same boxcar technique and window size mentioned above. Again, we subtracted the smoothed curve from the data and computed its residuals. The periodicity will be given by:
\begin{equation}
    Q = \frac{\sigma_{resid}^{2}-\sigma_{phot}^{2}}{\sigma_{raw}^{2}-\sigma_{phot}^{2}},
\end{equation}
where $\sigma_{resid}^{2}$ and $\sigma_{raw}^{2}$ are the variance of the residual and raw light curves, respectively. In our case, we choose $\sigma_{phot}$ to be the median of the photometric error of each light curve.

\begin{figure}
	\includegraphics[width=\columnwidth]{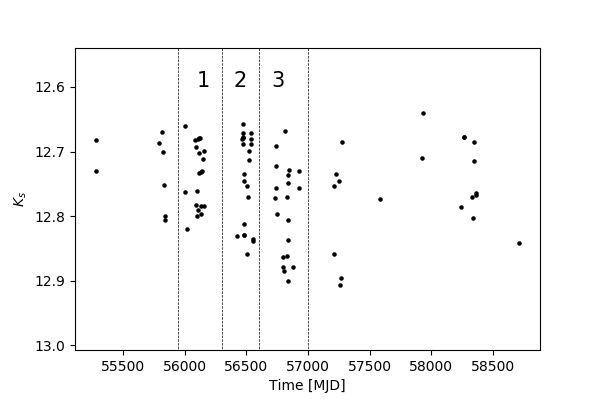}
    \caption{Light curve example for a given source in our sample. Dashed vertical lines are delimiting the three seasons with most consecutive data points (labeled by numbers). Period was computed for each of the three and then compared to the period obtained with the entire baseline of observation.}
    \label{fig:lc_example}
\end{figure}

\begin{figure}
	\includegraphics[width=0.95\columnwidth]{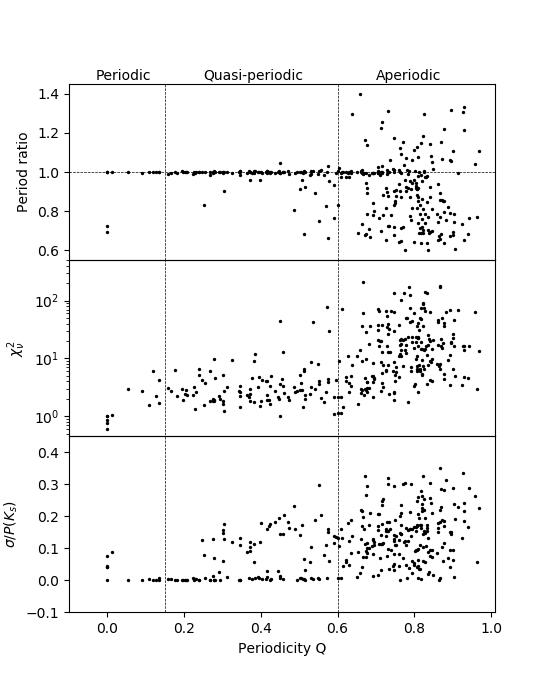}
    \caption{\emph{Upper panel:} Period ratio vs. periodicity $Q$ for sources in our sample. The horizontal dashed line corresponds to period ratios equal to $1$. \emph{Middle panel:} $\chi^{2}$ with respect to the smoothed light curve normalized by the number of data points of each light curve $\chi^{2}_{\nu}$, in logarithmic scale, against the periodicity $Q$ of sources in our catalog. A higher dispersion of $\chi^{2}_{\nu}$ is observed for aperiodic sources. \emph{Lower panel:} Standard deviation of the periods for each higher cadence season over the period of the star $\sigma / P(K_{s})$ against $Q$. The three panels only show sources with P$\leq40$ days and dashed vertical lines indicate the different limits for each periodicity classification. }
    \label{fig:Pratio_ChiQ}
\end{figure}

Due to the uneven cadence of VVVX data, it is difficult to detect by eye an appropriate limit between periodic and quasi-periodic, as it is difficult to see whether or not it has a pattern that repeats over time. Therefore, first we checked if the periods of the sources in our sample remained stable. To do so, we defined three observing seasons with relatively higher cadence, as those between $56000$ and $57000$ MJD (see Fig.~\ref{fig:lc_example}). For each star, we computed the periods in each of the three seasons, restricting the search close to the period $P(K_{s})$ based on the full time baseline. Then, we defined a period ratio, as the median of the three seasonal periods over the original $P(K_{s})$. The behaviour of this ratio against the periodicity parameter $Q$ is shown in the upper panel of  Fig.~\ref{fig:Pratio_ChiQ}. In this plot, the vast majority of the periods remain fairly stable for low values of $Q$. Only above $Q=0.6$ the larger dispersion indicates that the periods are not stable, so we classify these sources as aperiodic.

In order to account for changes in the shapes of the light curves, we computed the $\chi^{2}_{\nu}$ value for all the sources comparing the raw data to their smoothed version obtained to measure the periodicity parameter. Low values of this metric are expected for sources whose smoothed light curve is similar to the folded one. On the contrary, high $\chi^{2}_{\nu}$ describe curves with larger dispersion with respect to their smoothed version. The dependence of this parameter against the periodicity is shown in the middle panel of Fig.~\ref{fig:Pratio_ChiQ}, where aperiodic sources show higher $\chi^{2}_{\nu}$ values and dispersion, confirming the limits imposed for each periodicity classification. Periodic and quasi-periodic sources, on the other hand, have similar $\chi^{2}_{\nu}$, most likely because we hit the limit of the precision of VVVX data. 

Having the periods for each season of higher cadence, we computed their standard deviation $\sigma (K_{s})$ and compared it with the period obtained through all the baseline $P(K_{s})$. This, mainly because we found a number of points in the aperiodic region in which the period ratio was near $1$. The lower panel of Fig.~\ref{fig:Pratio_ChiQ} shows $\sigma/P(K_{s})$ against the periodicity. In the aperiodic part of this plot, we can observe that they are not actual stable periods because $\sigma/P(K_{s})$ is highly dispersed. Overall, we adopted the same periodicity limits for each category as \citet{Cody_2014}: periodic light curves have $Q\leq 0.15$, quasi-periodic have $0.15<Q<0.6$, and the others are aperiodic.

\begin{figure}
	\includegraphics[width=0.95\columnwidth]{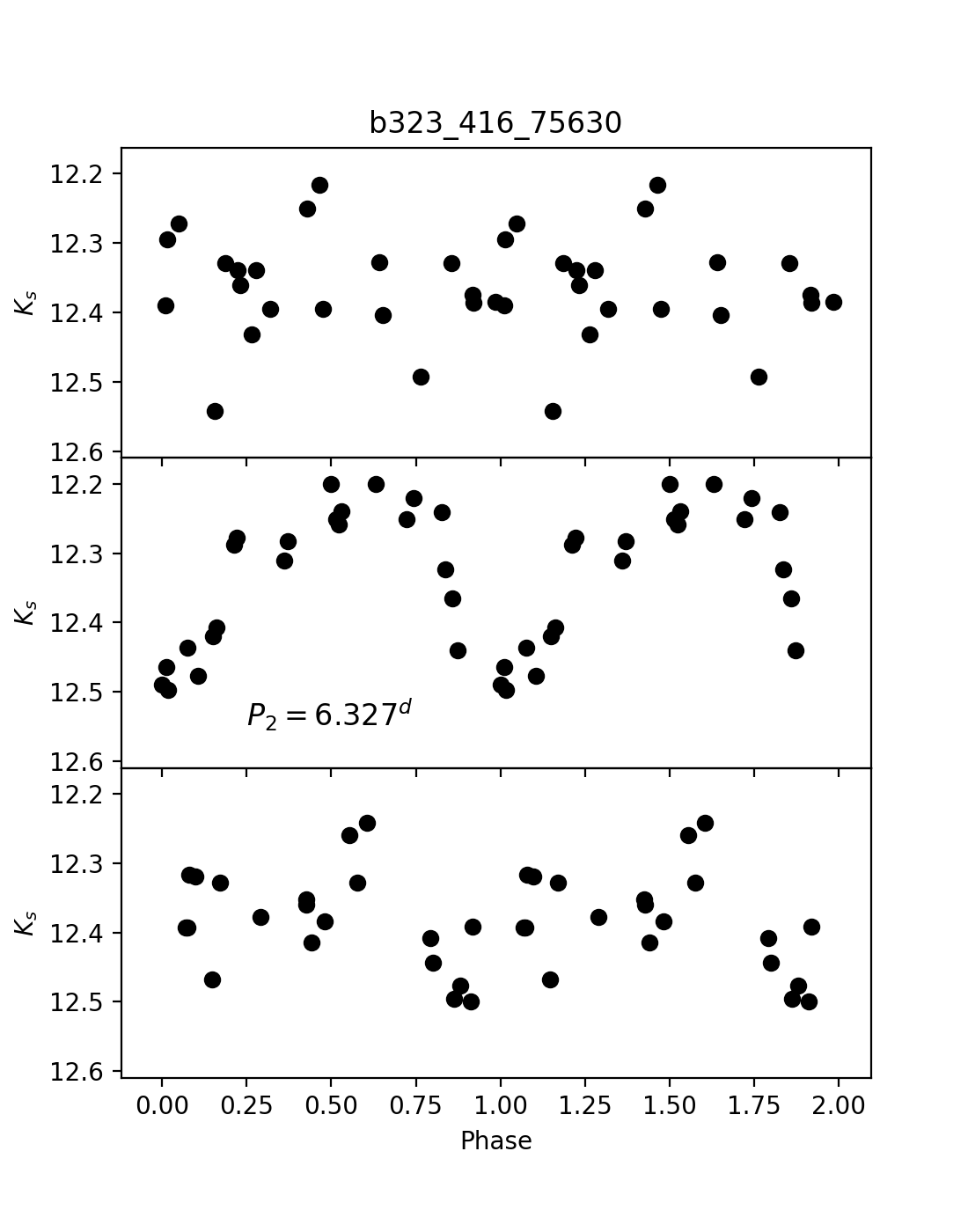}
    \caption{Example of a phased light curve for a star classified as aperiodic using the entire baseline of observation. From top to bottom, each panel represents the three seasons $1-3$ with higher cadence. In season $2$ (middle panel) the star showed signs of being periodic. In the three panels we have folded the light curves according to the period value in season $2$, $P_2$, which is indicated in the plot.}
    \label{fig:shape_change_aperiodic}
\end{figure}

Finally, it is important to mention that we are using the entire set of data points to compute a period and, therefore, the periodicity metric. This means that, if a star showed a periodic variability behaviour in just a part of the baseline, it will be classified as aperiodic. This effect would be larger for stars with higher photometric error values. In order to quantify the incidence of this in our data, we plotted the phased light curves of stars categorized as aperiodic for each of the three seasons with more cadence, already mentioned in Fig.~\ref{fig:lc_example}, in order to look for signs of noticeable periodic variability by eye in, at least, one of those seasons. We found that only $17$ out of $203$ aperiodic stars changed their periodicity behaviour. An example of this periodicity mutation can be observed in Fig.~\ref{fig:shape_change_aperiodic}, where, from top to bottom, seasons $1-3$ are located. Here, we can see that in season $2$ (middle panel) the variation of the light curve is periodic. However, we are keeping the classifications made through the complete set of data points.

\subsection{Asymmetry}

The asymmetry parameter, $M$, expresses the tendency of a light curve to increase or decrease its brightness. It also separates the data into three classes defined as: \textit{bursting}, for light curves which increase in flux; \textit{dipping}, those with occasional decreases below a more or less quiescent state, and \textit{symmetric}, as those not showing a tendency to either brighten or fade. To quantify this behavior, this metric is defined as:
\begin{equation}
    M = \frac{\langle m_{10\%} \rangle - m_{raw}}{\sigma},
\end{equation}
where $\langle m_{10\%} \rangle$ is the mean of the $10\%$ of the brightest and dimmest points of a given light curve, $m_{raw}$ is the median of the original light curve, whereas $\sigma$ is the standard deviation of the data. Due to the sparse sampling of VVVX data, it is hard to observe short bursts or dips. Therefore, this metric only gave us a lower limit of what could actually occur. We considered, again, only the data points in the three seasons with higher cadence, shown in Fig.~\ref{fig:lc_example} to compute the $M$ metric. This, because in these seasons, the probability of observing a real dip or burst is higher. 

The limits for the classification using this parameter are different from the ones proposed by \citet{Cody_2014}. In our study, symmetric sources are the ones with values $-0.40\leq M \leq 0.40$, whereas dipping ones have $M>0.40$ and bursting are the ones with $M<-0.40$. These limits were imposed because when ordering them from lowest to highest $M$ values, the dips or bursts were mostly observed in those ranges.

\begin{figure*}
	\centering
	\includegraphics[scale=0.46]{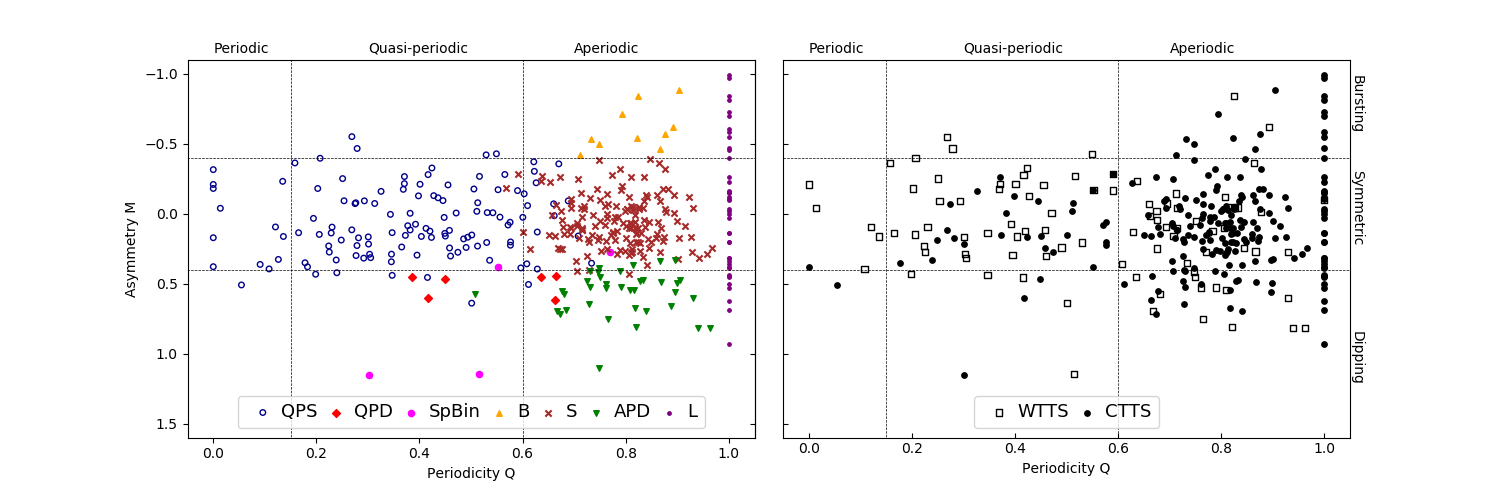}
    \caption{\emph{Left:} $Q-M$ plane for the $379$ sources in our catalog, divided into nine different regions according to their degree of periodicity and asymmetry. Colors represent our classifications by eye. (QPS: quasi-periodic symmetric; QPD: quasi-periodic dipper; SpBin: spectroscopic binary; B: burster; S: stochastic; APD: aperiodic dipper; L: long timescale). \emph{Right:} Same as the left panel, but now markers show the T Tauri status of our sources, when available. Open squares are WTTSs and black dots represent CTTSs.}
    \label{fig:QMplane}
\end{figure*}

In left panel of Fig.~\ref{fig:QMplane} we show the $Q-M$ plane, together with the limits defined above, between each class. The plot is divided into nine regions in which only six of them are populated or match our classifications by eye. This last was done by visually inspecting each light curve and identifying whether the classification given by the numerical values of the metrics alone agreed with what was observed in each of the light curves. Following the same notation of \citet{Cody_2014}, quasi-periodic symmetric are QPS, QPD is for quasi-periodic dipping ones, whereas aperiodic and bursting are denoted as B (bursters). Aperiodic and symmetric light curves are marked as S, from stochastic, and aperiodic dipping ones are tagged as APD. Long timescale sources are denoted as L. It is important to mention that here, we did not consider the periodic class, due to the fact that we did not observe a significant difference between them and the quasi-periodic light curves. In addition, we labeled as SpBin the spectroscopic binaries that were introduced in the study of \citet{Henderson_2012}. 

We found that, in a few cases, the $Q-M$ metric did not reflect what we observed in the light curve, and therefore Fig.~\ref{fig:QMplane} (left) also shows the sources color coded according to our visual classification. 

Note that some sources had negative $Q$ values, although with periods identical to the ones reported in \citet{Henderson_2012}. We kept those sources in the sample, but fixed their periodicity as $Q=0$. Negative values are non-physical for this parameter and, in our case, this is mainly a consequence of the amplitude of variation being comparable to the median of the error in the light curve. In addition and as mentioned before, long timescale sources have $Q$ values that do not represent what we observe on their light curves. Many of them are classified as quasi-periodic with periods greater than $40$ days. In these cases, we set their periodicity to $Q=1$. 

All the stars included in our catalog, along with their coordinates, $K_{s}$ magnitudes and classifications are listed on Table \ref{tab:tts} from Appendix \ref{sec:appendix1}. In addition, although stars with artificial periods were not included in our catalog, $40$ of them passed our filters in terms of data points and true variability. We list them on Table \ref{tab:alias} from Appendix \ref{sec:appendix2}, along with the same parameters mentioned above, and including their literature period, when available. However, we did not present their $Q$ metric, as it relies on the period value.

We show the classification of T Tauri stars in the right panel of Fig.~\ref{fig:QMplane}. Overall, most of the sources are aperiodic, few sources are classified either as periodic or quasi-periodic. The latter are stars with a stable period, regardless of the evolving shape of their curves. According to our visual inspection, we could not observe a marked difference between light curves classified as periodic and the ones categorized as quasi-periodic. 

\begin{figure}
	\includegraphics[width=\columnwidth]{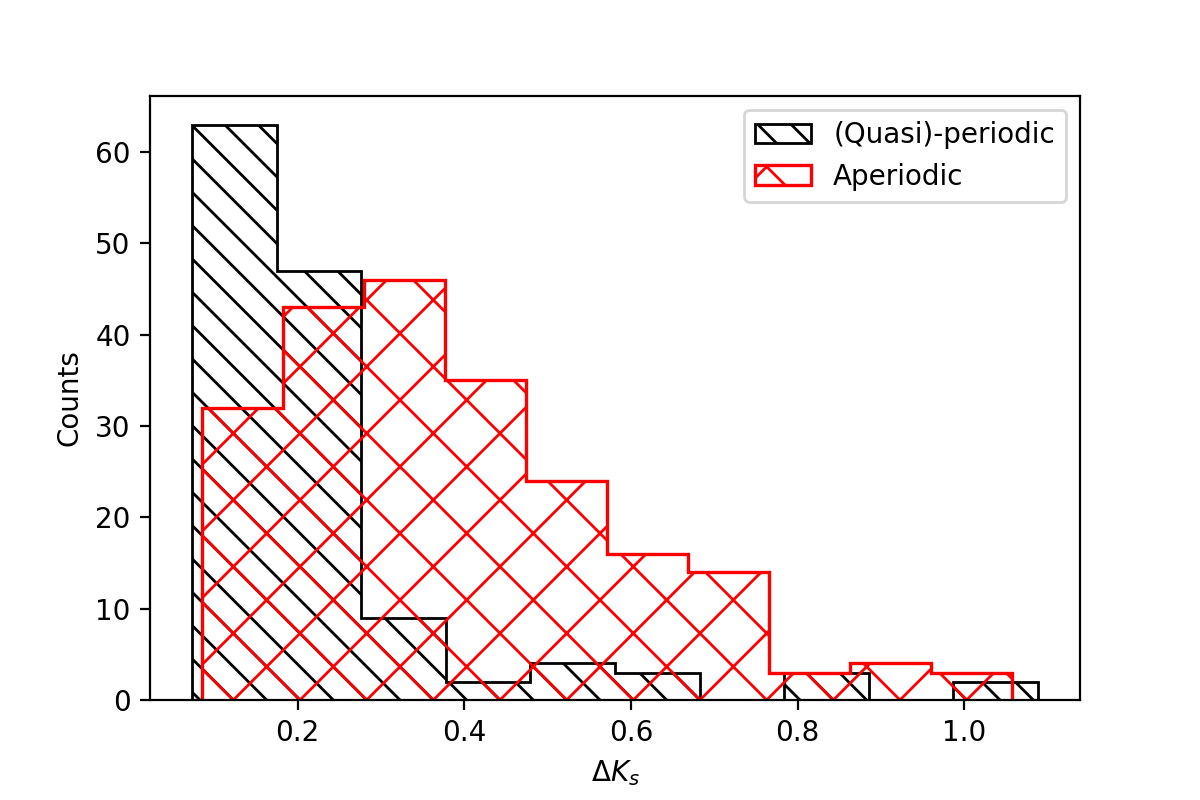}
    \caption{Amplitude distribution $\Delta K_{s}=K_{s,max}-K_{s,min}$ for stars in our catalog categorized as (quasi)-periodic (in black) and aperiodic (red).}
    \label{fig:amplitude}
\end{figure}

The amplitude distribution, $\Delta K_{s}=K_{s,max}-K_{s,min}$, for (quasi)-periodic and aperiodic stars is shown in Fig.~\ref{fig:amplitude}. Most of the (quasi)-periodic ones have amplitudes around $0.2$ mag, consistent with variations due to spots. Aperiodic sources have a more extended distribution of amplitudes. 

In fact, most aperiodic variables are CTTSs, but this is due to our catalogue being biased in favor of this kind of stars. Only a fraction of CTTSs have a stable period across the whole time baseline, as natural for stars accreting from a disk  \citep{Stauffer_2016}. 

WTTSs, on the contrary, are mostly located in the (quasi)-periodic region of the plot, with only a few of them lacking a constant period. We believe that this latter circumstances is likely due to the fact that variability due to spots has small amplitude, especially at long wavelengths, which in our case can be comparable to the photometric errors. Precisely for this reason, we did not require the errors to be smaller than 10$\%$ of the amplitude for the WTTSs for which we recovered the (rotation) periods quoted in \citet{Henderson_2012}.

\begin{figure}
	\includegraphics[width=\columnwidth]{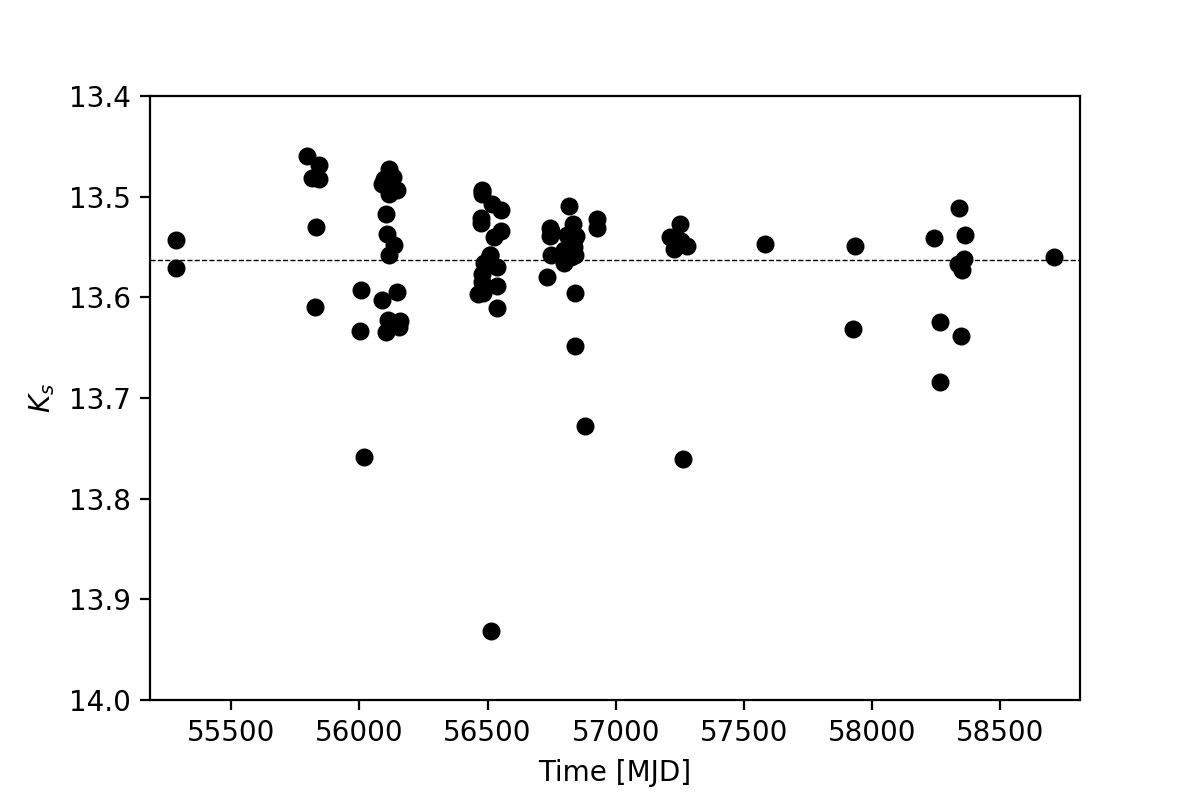}
    \caption{Light curve of a WTTS classified as APD. The dashed horizontal line indicates the mean of the points, related to a near constant level of flux. Points that deviate substantially from the mean are indicating the dips.}
    \label{fig:wtts_dipper}
\end{figure}

In addition, a few WTTSs appeared either as burster or dipper, as shown in Fig.~\ref{fig:QMplane}. Bursts in these stars could be attributed to flares, which do occur in these sources. Actually, just $2$ of them exhibited a burst. However, dips are not expected for stars without disks  (c.f., Fig.~\ref{fig:wtts_dipper}). The $15$ dips observed in the $93$ WTTSs of our catalogue could be due either to the presence of a binary companion, or due to disks with large inner radii, or to a combination of both. In the case of a binary, eclipses will be observed as dips in the light curve. If the stars do have disks, but with a large inner radius, they may or may not show IR excess, depending on how far the disk is \citep{Morales_Calder_n_2011}. If the stars are quasi-periodic, they are more likely to have binary companions.

Concerning asymmetry, the majority of the sources are symmetric, with values $-0.40 \leq M \leq 0.40$. Only a few sources have bursts in their light curves. This is partly due to the sparse cadence of VVVX preventing us to observe bursts of just a few hours. In addition, the effect of bursts is smaller at longer wavelengths because, depending on the viewing angle, the near IR emission
could be dominated by the disk, which prevents us to observe the bursts. The sources with long timescales variations are almost entirely CTTSs. Therefore, these changes could be due to changes in the luminosity or shape of the disk \citep{Cody_2014}, particularly in its outer parts. Furthermore, accretion can also impact the variability at this timescale. 

\begin{figure}
	\includegraphics[width=\columnwidth]{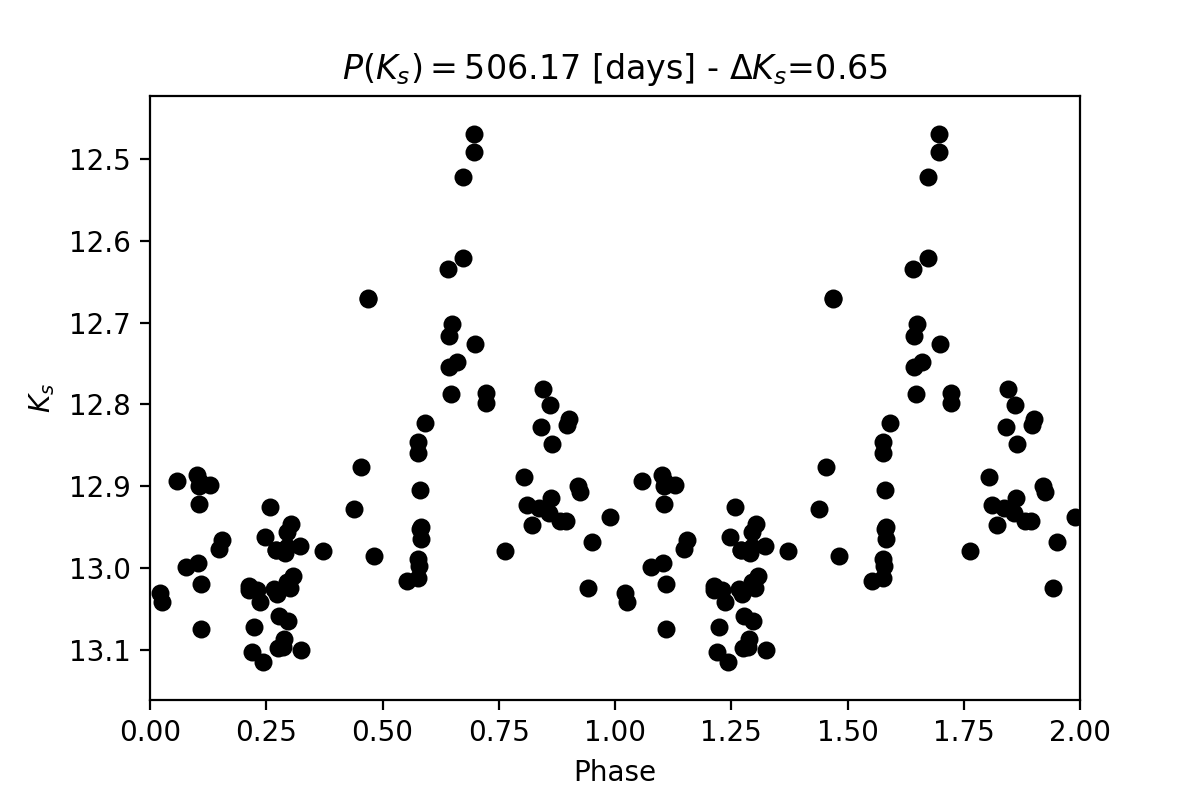}
    \caption{Folded light curve of a long timescale source in our catalog (ID b$323\_616\_15784$ in our catalog). The star showed repeated bursts with a timescale of $P=506.17$ days and may be a periodically outbursting YSO.}
    \label{fig:outburst}
\end{figure}

Interestingly, one of the sources in our catalog (ID b$323\_616\_15784$) showed periodic outbursts, very similar to the ones described in the literature (see, for instance, \citet{Dahm_2020, Guo_2022}). The timescale of its variation is around $500$ days and its folded light curve is shown in Fig.~\ref{fig:outburst}. The curve shows a fast rise followed by a slow decay. The shape of the light curve resembles the one of a periodic outburster YSOs, with an amplitude of variation $\Delta K_{s}=0.65$.

\begin{figure*}
	\centering
	\includegraphics[scale=0.66]{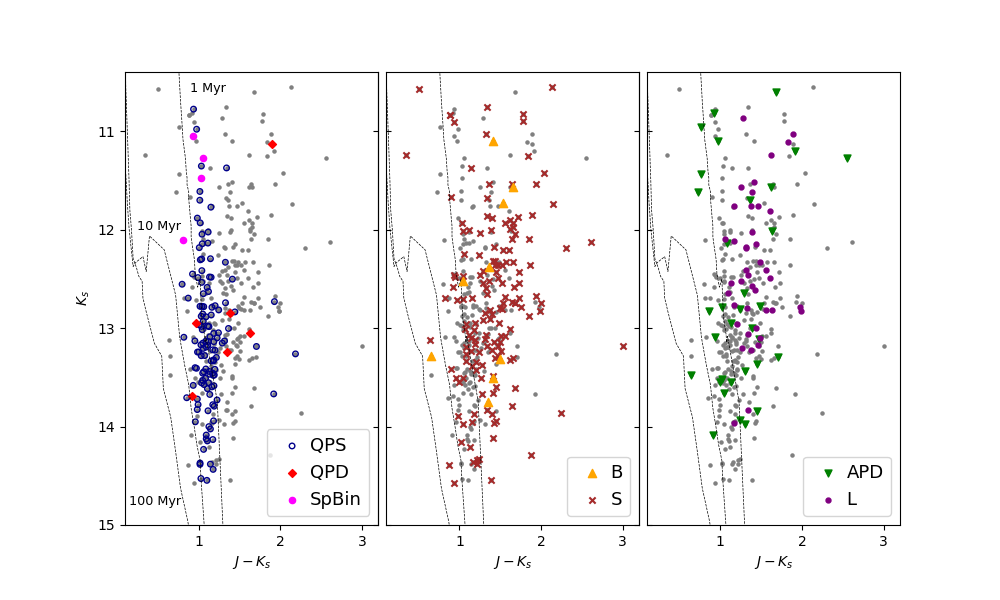}
    \caption{CMD for the T Tauri stars in our catalog (gray dots), coloured by their $Q-M$ classifications. These are shown in different panels just to avoid overlap and highlight how the different categories populate the CMD. Isochrones from \citet{Siess_2000} are also shown for the ages indicated in the left panel.}
    \label{fig:CMD_QM}
\end{figure*}

Finally, Fig.~\ref{fig:CMD_QM} shows the CMD for the sources in our catalog color coded by their classification according to the $Q$ and $M$ metrics. The majority of QPS stars are located in the region of less reddening, confirming that their flux changes are mainly related to spots. Some QPD sources are also in this region, which allow us to infer that the dips could be more likely related to eclipses of binary companions. On the other hand, QPD stars in the red part of the plot could indicate that their dips are related to the presence of a disk. The same occurs to the ones classified as B, S, APD and L, which also likely vary their brightness due to the presence of this structure. Accordingly, the position of stars in the CMD confirms in which region of the star the brightness variations originate.\\

\section{Comparison with previously found periods}
\label{sec:comparison_periods}
Forming stars are still contracting as they approach the MS. Due to the conservation of angular momentum, they are expected to speed up, even reaching velocities near the breakup. Nevertheless, it has been observed that stars usually rotate much more slowly than this value \citep{StaufferHartmann1987}. One of the explanations for this low stellar rotation rates comes from the disk-locking effect. Stars surrounded by disks are expected to rotate more slowly than the ones without such structures. 

Moreover, it is expected that for a viscous accretion disk, a transfer of angular momentum from the inner to the outer disk takes place. This, combined with accretion itself leads to a disk braking, slowing down the star rotation. This process, however, is not instantaneous: the time it takes for the star to slow down depends on its accretion rate \citep{Hartmann_2002}.

Consequently, accretion plays a role in the evolution of the angular momentum of stars \citep{Koenigl_1991} and this can be studied by measuring the rotation velocity of stars of several masses and in different evolutionary phases. The presence of spots in the photosphere helps measuring this velocity. If spots remain stable across the entire baseline of the observations, they imprint a sinusoidal variation in the light curves, allowing us to measure the rotation period. 

\begin{figure}
	\includegraphics[width=\columnwidth]{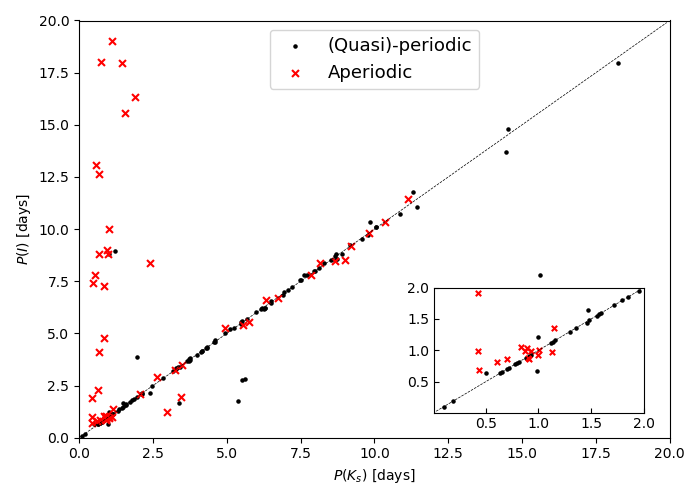}
    \caption{Period comparison between the values found in \citet{Henderson_2012} $P(I)$ and the ones computed in this study $P(K_{s})$. Periodic or quasi-periodic sources are in black, while red crosses are representing stars having aperiodic VVVX light curves. The gray dashed line indicates the one-to-one relation. The inset panel shows a zoom to period values less than $2$ days in both bands.}
    \label{fig:Percomp_IKs}
\end{figure}

\citet{Henderson_2012} obtained the rotation periods for $290$ low-mass pre-MS stars in the Lagoon Nebula region. They analysed photometric data collected for about $35$ nights (June-July, 2006) to build I-band light curves, which was obtained over a non-overlapping time range with the VVVX data. Out of the $174$ stars in common with their sample, for $126$ we found identical periods (see Fig.~\ref{fig:Percomp_IKs}). The vast majority of the common sources are classified either as periodic or quasi-periodic (black dots) in VVVX data and almost all of them lie in the one-to-one relation (dashed line).

However, for some sources, classified as aperiodic in this work (red crosses), the peaks of the periodogram are superimposed on a higher level of noise. These light curves have an amplitude of variation comparable to the median of their errors, leading to higher values of the $Q$ metric. So that, even if the period is similar to the one in the I-band, its $Q$-value will be large.

There are also common sources for which we could not find a stable period, although they were periodic in the I-band observations. Because the amplitudes of spot modulated light curves are expected to decrease as the wavelength of observation increases, the difference between the periodicity classification could be due to low amplitudes in the I-band that are not measurable in the $K_{s}$-band \citep{stahler}. Let us also keep in mind that VVVX light curves are noisier than the ones in \citet{Henderson_2012}.

\begin{figure}
	\includegraphics[width=\columnwidth]{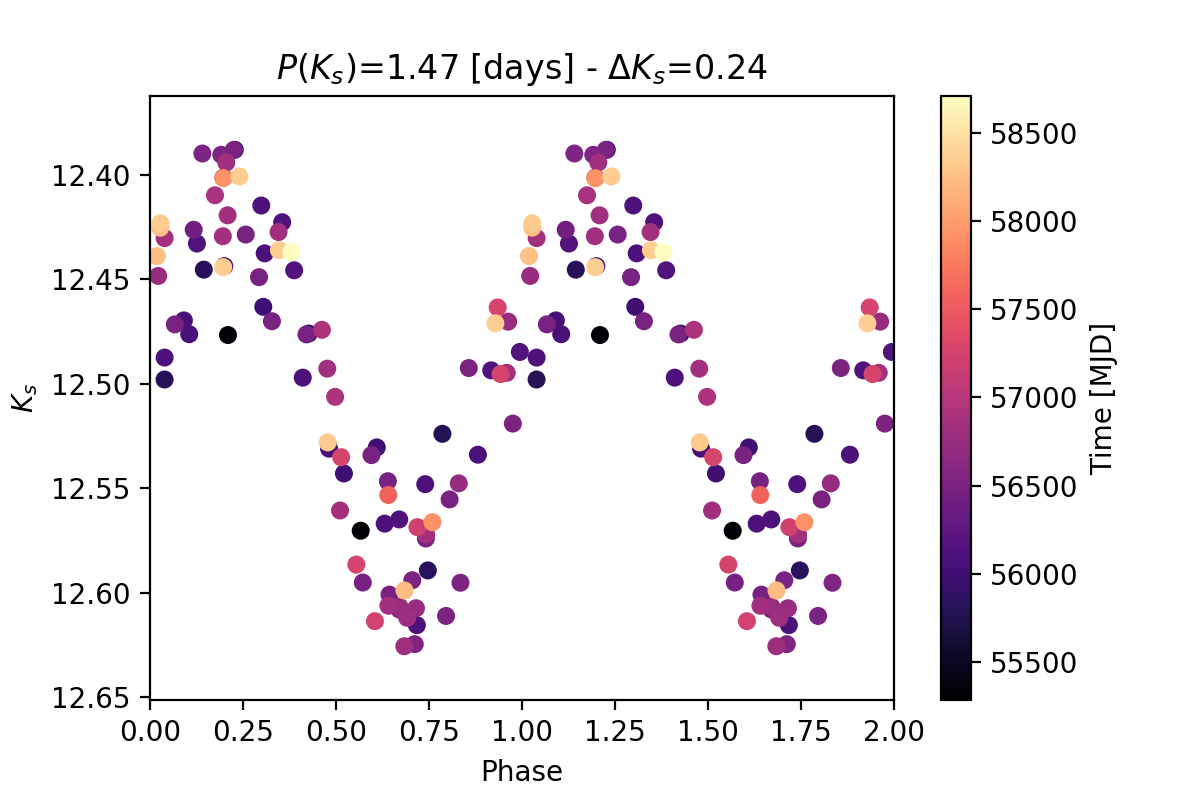}
    \caption{Phase folded light curve of a star classified as periodic (P) in our catalog. Data points are coloured according to the time of observation.}
    \label{fig:quasi_periodic}
\end{figure}

It is anyway remarkable that for most of the stars, the periods remain highly stable for years. This can be observed in Fig.~\ref{fig:quasi_periodic}, where we have plotted the folded light curve for a star classified as periodic. Epochs or data points are coloured by the time of observation. This result confirms the one found in \citet{Grankin_2008}, but in our case, the sample of stars is larger.
Furthermore, the mean of the fractional difference between periods in the I-band and the ones in the near-IR, defined as $(\lvert (P(I)-P(K_{s})) \rvert / P(K_{s}))$, has a value of $0.9 \%$, and the mean amplitude of the stars in common is $\Delta K_{s}\sim0.2$ magnitudes, which is expected for variations due to cool spots. For these stars, as their near-IR variability correlates with the one in the optical, we can conclude that it has a common origin: the presence of spots in the stellar photosphere. 

\begin{figure}
	\includegraphics[width=\columnwidth]{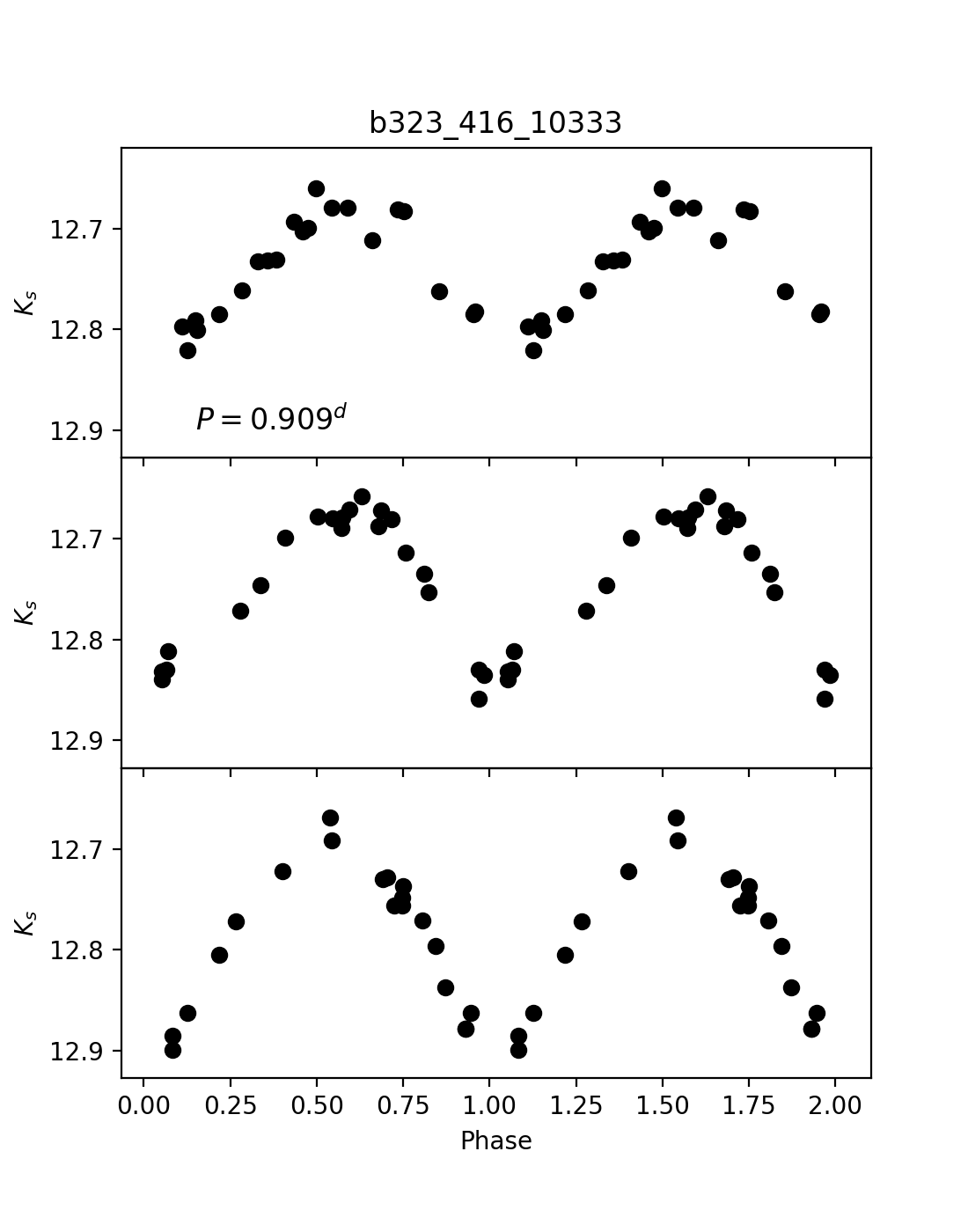}
    \caption{Phase folded light curve of a star that has similar periods both in the I- and $K_s$-band. Each panel has data points belonging to seasons $1-3$ with most consecutive measurements, from top to bottom. The period, indicated in the top panel, remains stable in all three seasons.}
    \label{fig:shape_change}
\end{figure}

Furthermore, due to differential rotation of the stars, this means that these structures remain at the same latitude for years. Thus, magnetically active regions on the stellar surface are located in a preferential area of the star. However, within eight years of observation, spots are evolving, changing their sizes or temperatures, and this modifies the shape of the light curve that we observe \citep{Grankin_2008}, leading to a quasi-periodic variable behaviour. An example of the evolving pattern of a given star can be observed in Fig.~\ref{fig:shape_change}. In this plot we show the phased light curve for a star which has the same period value both in the I- and $K_s$-band. Each panel represents the three seasons $1-3$ with higher cadence (as in Fig.~\ref{fig:lc_example}), from top to bottom. One can observe, from one season to another, that the shape of the light curve is not strictly the same, the amplitude of the variation is changing, but the phase remains fairly stable. We verified by visual inspection that this also happens for the great majority of stars classified as (quasi)-periodic. 

From the aforementioned we interpret that patterns and amplitude variations come from individually changing spots. However, the region where they develop remains magnetically active for, at least, several years. 

On the other hand, as we also have sources that are classified as aperiodic, even in cases where the period is similar to the one in the I-band, we found that $10$ of those stars showed signs of being periodic in one of the seasons with higher cadence. In fact, the star shown in Fig.~\ref{fig:shape_change_aperiodic} is one we have in common with the study of \citet{Henderson_2012} and presented similar period values in both bands ($P(I)=6.590^{d}$ and $P(K_s)=6.327^{d}$), but was classified as aperiodic. 

In addition, from the study of \citet{Prisinzano2019}, we have information on the temperatures of several sources and the majority of the ones that we include in our sample are of K spectral type. Spot lifetimes are expected to depend on temperature \citep{Basri_2022} and to be longest at these temperatures.

\section{Correlation with optical variability}
\label{sec:correlation_optical}
Brightness changes in PMS stars are a panchromatic phenomenon. Observations at different wavelengths provide information at different spatial scales with respect to the center of the star \citep{Venuti_2021} which, in turn, are related to different physical phenomena. Optical photometric variability is mostly related to the presence of spots and/or variable extinction. On the other hand, near-IR fluxes are less affected by spots and variable extinction, but are sensitive to structural changes in the disk of the star \citep{Chelli_1999, Bertout_2000}. If correlation exists between the changes in optical and near-IR, it indicates that the variability has a common origin, although the details depend on the viewing angle, the disk flux, its geometry, and the number of processes causing the variability at the same time \citep{Cody_2014}. On the other hand, when no correlation exists, it indicates that it originates from different regions of the star, such as the photosphere or the disk \citep{Eiroa_2002}. 

In the study of \citet{Venuti_2021}, $278$ members of the Lagoon Nebula region, from B to K spectral types, were studied. Using Kepler/K2 data of Campaign 9 \citep{Borucki_2010, Howell_2014}, they monitored this region for approximately two months. The high cadence of Kepler ($\sim 30$ minutes) allowed the authors to classify the light curves according to their degrees of periodicity and asymmetry. 

\begin{figure}
	\includegraphics[width=\columnwidth]{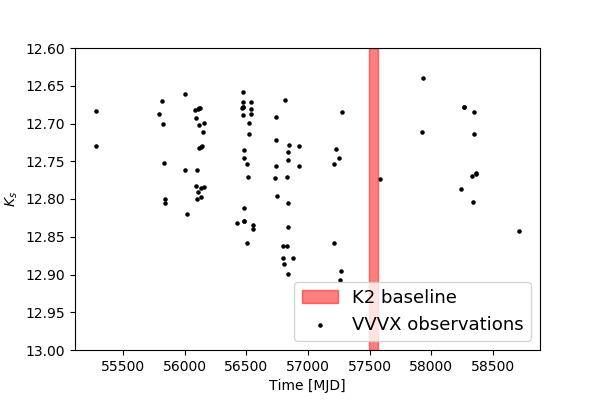}
    \caption{Example of a VVVX light curve, where observations span roughly eight years, compared to the roughly two months of Kepler/K2 monitoring (red strip).}
    \label{fig:Baseline_comparison}
\end{figure}

To check if there are correlations between our classifications in the $K_{s}$-band and the ones in the optical, we cross-referenced the VVVX and K2 data, finding $55$ common stars. Both baselines overlap in a certain time range, as can be seen in the light curve example from Fig.~\ref{fig:Baseline_comparison}, where the red strip marks the observation window of K2. However, they are not exactly simultaneous. Due to the different baselines of observations, we only compared stars with periods or timescales similar to the ones in the Kepler band, i.e., less than $40$ days in the VVVX data. Also, as we did not include YSOs of early spectral type, the comparison was made for F-type stars and cooler, resulting in a not very large number of sources in common.

\begin{figure*}
	\centering
	\includegraphics[scale=0.46]{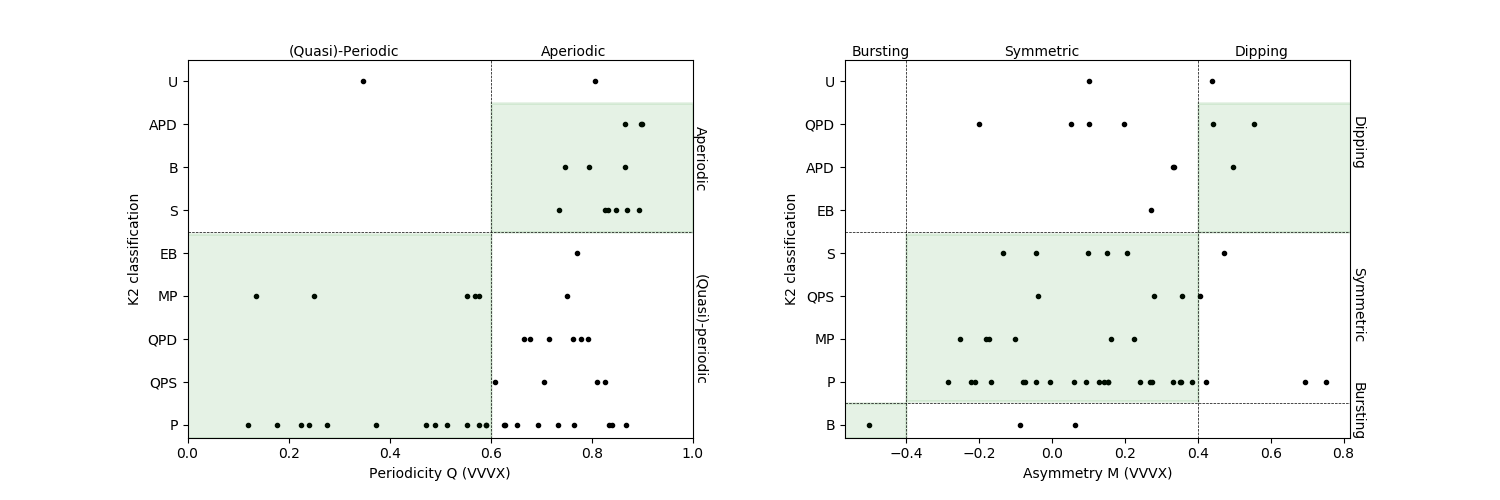}
    \caption{\emph{Left panel:}Kepler/K2 classification compared to the periodicity values found for VVVX light curves. Green shaded regions represent parts of the plot where classifications are either periodic or (quasi)-periodic in both bands. \emph{Right panel:} Kepler/K2 classification compared to the asymmetry values found for VVVX light curves. Green shaded regions represent parts of the plot where classifications are similar in both bands (symmetric, dipping or bursting). Classifications by eye in VVVX data are not shown in these plots.}
    \label{fig:QM_K2_VVVX}
\end{figure*}

The classification comparison for the common stars is shown in Fig.~\ref{fig:QM_K2_VVVX}, for the periodicity (left panel) and asymmetry (right panel). For each, we compared their $Q$ and $M$ values in the $K_{s}$-band with their classifications in the optical. Based on their $Q$ value, $33$ sources appeared with a stable period in both bands. They fall in the green shaded regions of Fig.~\ref{fig:QM_K2_VVVX}. According to our visual inspection, $6$ additional stars were classified as quasi periodic. This gave us a total of $39$ out of $55$ stars with consistent periodicity. We recall that, by eye, we could not observe a marked difference between VVVX light curves classified as periodic with the quasi-periodic ones. Moreover, the vast majority of the stars classified as periodic in the optical have spots, which can be stable for months but usually change in a time span of years, due to a variation in size, temperature or location within the magnetically active region \citep{Grankin_2008}, leading to a quasi-periodic variation.

For stars with different periodicity classification, we attribute this to three possibilities. For stars without disks and periodic variations due to spots, if this is not observed in the $K_{s}$-band, it could be related to the smaller amplitude of variation at this wavelength. Nevertheless, if a star does have a disk, it could appear as aperiodic in the near-IR because we would be observing reprocessed light from the disk and its variable extinction. Furthermore, if a star showed a periodic behavior in only a fraction of the baseline, it could be classified as aperiodic in the $K_s$-band.

For the asymmetry comparison, right panel of Fig.~\ref{fig:QM_K2_VVVX} shows that $34$ stars are classified as symmetric in both bands. However, some dips or bursts observed in the K2 data are not noticeable in the near-IR, most likely due to the short duration of these events, combined to the sparse cadence and lower precision of VVVX data. Due to these we are only sensitive to observing a lower limit for these type of events. The only burst that we were able to detect was the one with the largest amplitude. In summary, there are $39$ out of $55$ stars with consistent asymmetry classification, either by the $M$ value or by our visual inspection. There are also $5$ sources that were classified as dippers in the near-IR, but appeared as symmetric in the optical. Based on our visual inspection, $2$ of these stars had actual symmetric variations. The other $3$ showed dips in their $K_{s}$-band light curves, but $2$ of them are near the saturation limit of VVVX data, so we can confidently state that only $1$ of them (a WTTS) showed real dips. As mentioned before, dips are not generally expected in these stars, but could be due either to the presence of a binary or to a disk with a large inner hole \citep{Morales_Calder_n_2011}.

Classifications in the work by \citet{Venuti_2021} also included a class named "unclassifiable" (U), which corresponded to light curves that did not match the categories defined above. We found that $2$ unclassifiable stars in the optical had a counterpart in our data. One of them was classified as stochastic and the other as quasi-periodic symmetric, considering our visual inspection. This last had an amplitude consistent with variations due to spots.

Overall, the classifications according to the $Q$ and $M$ parameters are in good agreement between both bands. In these cases, the brightness changes have a common origin. Otherwise, the changes originate from different regions of the star.

\section{Conclusions}
\label{sec:conclusions}
In this study, we have compiled a catalog of VVVX $K_s$-band light curves for $379$ T Tauri stars in the Lagoon Nebula region. The $8$ years baseline of VVVX allowed us, for the first time, to study near IR flux changes at different timescales, starting from a fraction of a day to years, in this kind of stars. The light curves were classified according to their degree of periodicity $Q$ and asymmetry $M$, following the work by \citet{Cody_2014}, but with the limits slightly modified to adapt them to our ground-based data. These parameters allowed us to infer the physical processes that cause the observed flux changes.  In terms of the $Q$ parameter and by visually inspecting our catalog of light curves, we did not find a marked difference between periodic and quasi-periodic sources. Nevertheless, there was a strong contrast between quasi-periodic and aperiodic stars.

In the $K_{s}$-band, CTTSs are mostly classified as stochastic (S), with only a small fraction showing a (quasi)-periodic behavior. This is expected because, for stars with disks, accretion is stochastic by nature. In addition, more than one process can affect the variability at the same time, such as obscuration by circumstellar material, instabilities in the accretion disk, hot spots that evolve in time, to name a few. All these processes, separately or in combination, lead to stochastic brightness changes.

However, CTTSs could also show a periodic behavior related to accretion. The presence of a companion can induce recurrent accretion events. As shown by simulations, mass flow, through streams, in circumbinary gaps can lead to accretion in one or both components of the system \citep{Artymowicz_1996}. These systems can involve two stars that near pericenter induce a higher accretion rate \citep{Dunhill_2015}. Likewise, planetary companions in eccentric orbits can produce pulsed accretion onto the central star \citep{Teyssandier_2020}. In any of those cases, a periodic variability is developed, named as {\it periodic outbursts} \citep{Dahm_2020, Guo_2022}. In fact, we found one source that could be considered as a candidate of one of this class, mainly because of its repetitive bursts in timescales of about $500$ days and the consistent shape of its folded light curve.

For WTTSs, the majority appear as (quasi)-periodic, which is related to the presence of dark spots developed in the magnetically active regions of the stars. A few show dips in their light curves, a non-common behavior possibly related to the presence of binary companions or to disks with large inner holes, which introduce IR excesses at longer wavelengths.

The periods obtained in our study were compared to the ones obtained in \citet{Henderson_2012}. We found that, despite the very different baselines of observation (35 nights in the I-band \emph{vs.} 8 years in the $K_{s}$-band), the majority of the common stars had very similar periods with a mean of fractional difference of about $0.9 \%$. Flux variations in these cases are mostly related to the presence of spots, whereby the magnetically active region remains very stable, even for years, as shown by VVVX data. This confirms the result of \citet{Grankin_2008}, but for a larger sample of stars. Besides, individual spots are expected to evolve within this region, changing in size, temperature or location \citep{Grankin_2008}, leading to long term (quasi)-periodic variations. This, along with the cadence and errors of our data, is the likely reason why we do not see a marked difference between periodic and quasi-periodic stars.

Overall, due to the stability of the phase and periods, we have proven that the magnetically active regions where spots develop remain stable for several years.

Some of the sources in our catalog were also studied by \citet{Venuti_2021}, in the optical, for $\sim2$ months of Kepler observations. They classified their light curves using the $Q$ and $M$ metrics. We compared our classification with theirs, for the 55 common sources (with P$<40^d$). The agreement is generally very good, although some stars appeared as stochastic in the near-IR, whereas they were (quasi)-periodic in the optical. We ascribe this to the fact that the amplitude of the variation due to spots is expected to be smaller at longer wavelengths, and they could be lost in the larger errors of VVVX data. Besides, if the star showed periodic variations in only a part of the entire VVVX baseline, it could be classified as aperiodic. Regarding the asymmetry metric, we found that dips and burst events are difficult to observe in our data, especially those of small amplitudes and short duration. Nonetheless, there were cases in which dips and bursts were observed in our catalog and followed the same classification as in the optical.

Despite the fact that the \citet{Cody_2014} $Q$ and $M$ metrics were designed to work on spaced-based data, they have proven to be a good tool for classifying light curves obtained from ground-based data. In the future, we plan to use them for light curve classification of PMS stars in other star-forming regions observed by the VVVX survey.

\section*{Acknowledgements}

C.O.H. acknowledges the support from National Agency for Research and Development (ANID), Scholarship Program Doctorado Nacional 2018–21180315. This project was funded by ANID FONDECYT Regular 1191505, ANID Millennium Institute of Astrophysics (MAS) under grant ICN12\_009, the ANID BASAL Center for Astrophysics and Associated Technologies (CATA) through grants AFB170002, ACE210002 and FB210003, and ANID, -- Millennium Science Initiative Program -- NCN19\_171. In addition, this project was partially funded by the Max Planck Society through a “Partner Group” grant. 

We gratefully acknowledge the use of data from the VVV ESO Public Survey program ID 179.B-2002 taken with the VISTA telescope, and data products
from the Cambridge Astronomical Survey Unit (CASU). The VVV Survey data are made public at the ESO Archive.
Based on observations taken within the ESO VISTA Public Survey VVV, Program ID 179.B-2002. 

We also thank Laura Venuti, who provided us with Kepler/K2 data of sources that we had in common with her study in the M8 region.

This publication uses data generated via the Zooniverse.org platform, development of which is funded by generous support, including a Global Impact Award from Google, and by a grant from the Alfred P. Sloan Foundation.

\section*{Data Availability}

The VVV and VVVX data are publicly available at the ESO archive \url{http://archive.eso.org/cms.html}. Light curves, obtained through PSF photometry of VVVX data, have not yet been publicly released but are available on request to the first author.



\bibliographystyle{mnras}
\bibliography{example} 




\appendix

\section{Parameters of T Tauri stars in our catalog}
\label{sec:appendix1}
Table \ref{tab:tts} lists different physical parameters for the T Tauri stars in our catalog of light curves. The complete table can be downloaded from.

\begin{table*}

    \caption{List of T Tauri stars that compile the VVVX light curve catalog with their equatorial coordinates, the mean $K_{s}$ magnitude and its mean error $\overline{eK_{s}}$, the periods obtained, the amplitudes of variation $\Delta K_{s}=K_{s, max}-K_{s, min}$ without outliers, the metrics for the classification of the light curves, $Q$ and $M$, and our classification by visual inspection.}
    \label{tab:tts}
    \resizebox{\textwidth}{!}{%
    \begin{tabular}{|l|c|c|c|c|c|c|c|c|c|}
    \hline
        ID & RA [deg] & DEC [deg] & $\overline{K_{s}}$ & $\overline{eK_{s}}$ & $P(K_{s})$ [days] & $\Delta K_{s}$ & Q & M & Class \\ \hline
        b323\_416\_69746 & 271.0921 & -24.4296 & 11.03 & 0.02 & 0.1 & 0.16 & 0.63 & 0.13 & QPS \\
        b324\_304\_22054 & 271.0271 & -24.2948 & 13.86 & 0.02 & 0.19 & 0.56 & 0.05 & 0.51 & P \\
        b323\_416\_9995 & 270.9835 & -24.3208 & 13.5 & 0.02 & 0.65 & 0.18 & 0.37 & -0.27 & QPS \\
        b324\_304\_44498 & 271.0825 & -24.3017 & 13.6 & 0.03 & 0.44 & 0.29 & 0.9 & 0.47 & APD \\
        b323\_416\_44777 & 271.0236 & -24.4153 & 12.09 & 0.01 & 0.72 & 0.07 & 0.0 & -0.21 & P \\ \hline
    \end{tabular}%
    }
\textbf{Note}. A part of Table \ref{tab:tts} is shown here for guidance regarding its content and structure. The entire table is published in the machine-readable format.

\end{table*}

\section{Parameters of T Tauri stars with artificial periods}
\label{sec:appendix2}
Several T Tauri stars were not included in our catalog due to their artificial period values. Therefore, the analysis for the periodicity $Q$, considered for the ones with actual periods, was no longer valid in these cases. Still, $40$ of them showed real variability on their light curves and passed our filters regarding the number of data points and saturation limit. As the asymmetry $M$ is not obtained through the period, we also present its value here. These are listed on table \ref{tab:alias}.

\begin{table*}
	\caption[Short caption]{T Tauri stars with artificial period values,  having light curves with more than $40$ data points, true variability and were below the saturation limit. Columns are: equatorial coordinates, mean magnitude $\overline{K_{s}}$ and mean error $\overline{eK_{s}}$, the amplitude of variation $\Delta K_{s}$, asymmetry $M$ and their periods, when available.}
	\label{tab:alias}
\resizebox{0.98\textwidth}{!}{%
\begin{tabular}{|l|c|c|c|c|c|c|c|}
\hline
ID & RA [deg] & DEC [deg] & $\overline{K_{s}}$ & $\overline{eK_{s}}$ & $\Delta K_{s}$ & M & P(I) [days] \\ \hline
b323\_416\_35940 & 271.0226 & -24.3758 & 14.09 & 0.02 & 0.22 & -0.22 & 3.26329 \\
b323\_416\_62915 & 271.1066 & -24.3686 & 13.5 & 0.02 & 0.17 & 0.11 & 4.07912 \\
b323\_416\_62215 & 271.0988 & -24.3778 & 11.98 & 0.01 & 0.15 & 0.22 & 6.01169 \\
b324\_304\_55190 & 271.1161 & -24.2967 & 12.89 & 0.02 & 0.44 & 0.76 & 7.10204 \\
b324\_304\_57845 & 271.1105 & -24.3178 & 12.64 & 0.02 & 0.3 & -0.5 & 9.46939 \\
b323\_416\_65354 & 271.1082 & -24.3804 & 13.08 & 0.02 & 0.23 & 0.36 & 12.59194 \\
b323\_416\_49023 & 271.0694 & -24.36 & 11.22 & 0.02 & 0.39 & 0.44 & 18.03399 \\
b323\_616\_6648 & 271.1596 & -24.3882 & 12.96 & 0.03 & 0.59 & 0.07 & 18.03506 \\
b323\_416\_2126 & 270.9179 & -24.3889 & 10.77 & 0.02 & 0.18 & 0.75 & - \\
b323\_416\_38869 & 271.0523 & -24.3405 & 12.33 & 0.02 & 0.23 & 0.32 & - \\
b324\_304\_18068 & 271.0553 & -24.233 & 12.89 & 0.02 & 0.25 & -0.29 & - \\
b323\_416\_69391 & 271.0642 & -24.4739 & 12.95 & 0.01 & 0.28 & 0.13 & - \\
b323\_416\_60961 & 271.0762 & -24.4077 & 13.33 & 0.02 & 0.47 & -0.17 & - \\
b324\_304\_45094 & 271.0969 & -24.2813 & 13.01 & 0.02 & 0.85 & 0.52 & - \\
b323\_416\_72262 & 271.1284 & -24.3811 & 11.44 & 0.01 & 0.12 & -0.02 & - \\
b323\_416\_72139 & 271.1288 & -24.3799 & 12.78 & 0.02 & 0.16 & 0.03 & - \\
b324\_304\_66077 & 271.1932 & -24.2256 & 13.7 & 0.02 & 0.43 & 0.07 & - \\
b323\_616\_41784 & 271.2229 & -24.4327 & 10.85 & 0.02 & 0.24 & 0.29 & - \\
b324\_104\_24915 & 271.2846 & -24.2591 & 10.92 & 0.02 & 0.25 & 0.76 & - \\
b323\_614\_39786 & 270.6398 & -24.2486 & 12.69 & 0.02 & 0.25 & 0.11 & - \\
b323\_614\_56738 & 270.6731 & -24.2697 & 14.57 & 0.03 & 0.54 & 0.08 & - \\
b324\_304\_18460 & 271.0174 & -24.2953 & 12.73 & 0.02 & 0.24 & -0.09 & - \\
b324\_304\_68671 & 271.1765 & -24.2644 & 13.87 & 0.02 & 0.62 & 0.19 & - \\
b323\_416\_63256 & 271.1039 & -24.375 & 13.27 & 0.02 & 0.16 & -0.07 & 0.92858 \\
b324\_203\_92473 & 270.9888 & -24.2694 & 12.4 & 0.02 & 0.57 & 0.62 & 2.42001 \\
b323\_416\_48451 & 271.065 & -24.3644 & 13.07 & 0.03 & 0.32 & -0.09 & 2.80857 \\
b323\_416\_15373 & 270.9467 & -24.4059 & 11.68 & 0.01 & 0.41 & 0.16 & 4.07912 \\
b323\_416\_70575 & 271.104 & -24.4137 & 12.79 & 0.01 & 0.18 & -0.02 & 4.28333 \\
b323\_416\_59753 & 271.0546 & -24.4371 & 12.02 & 0.02 & 0.31 & 0.7 & 5.03891 \\
b324\_104\_45823 & 271.2403 & -24.4157 & 12.69 & 0.01 & 0.27 & -0.15 & 6.4654 \\
b324\_304\_33667 & 271.0561 & -24.297 & 14.09 & 0.02 & 0.2 & -0.13 & 8.30167 \\
b323\_416\_59018 & 271.0683 & -24.4108 & 11.19 & 0.02 & 0.47 & 0.6 & 8.56615 \\
b323\_416\_53019 & 271.0721 & -24.3744 & 12.22 & 0.02 & 0.36 & -0.09 & 8.78579 \\
b323\_416\_14766 & 270.9745 & -24.3577 & 13.47 & 0.02 & 0.33 & -0.1 & 9.72481 \\
b324\_304\_59950 & 271.1028 & -24.3406 & 12.05 & 0.01 & 0.28 & -0.17 & 11.7551 \\
b324\_304\_15754 & 271.0213 & -24.2784 & 12.21 & 0.02 & 0.24 & -0.72 & 12.18647 \\
b323\_416\_60310 & 271.095 & -24.3734 & 13.43 & 0.02 & 0.18 & 0.15 & 3.0869 \\
b324\_104\_37374 & 271.2151 & -24.421 & 11.57 & 0.02 & 0.22 & 0.38 & 3.6845828 \\
b323\_616\_8340 & 271.1165 & -24.4666 & 12.33 & 0.02 & 0.44 & 0.57 & 11.4222067 \\
b323\_416\_67173 & 271.0809 & -24.4353 & 12.92 & 0.02 & 0.51 & -0.32 & 14.8976478 \\ \hline
\end{tabular}%
}
\end{table*}


\bsp	
\label{lastpage}
\end{document}